\documentstyle[sprocl,epsf,rotating,axodraw]{article}

\catcode`@=11
\def\citer{\@ifnextchar
[{\@tempswatrue\@citexr}{\@tempswafalse\@citexr[]}}

\def\@citexr[#1]#2{\if@filesw\immediate\write\@auxout{\string\citation{#2}}\fi
  \def\@citea{}\@cite{\@for\@citeb:=#2\do
    {\@citea\def\@citea{-\penalty\@m}\@ifundefined
       {b@\@citeb}{{\bf ?}\@warning
       {Citation `\@citeb' on page \thepage \space undefined}}
	{\csname b@\@citeb\endcsname}}}{#1}}
\catcode`@=12

\newcommand{\st}{\tilde{t}}

\newcommand{\sq}{\tilde{q}}
\newcommand{\sqb}{\bar{\tilde{q}}}
\newcommand{\gl}{\tilde{g}}
\newcommand{\gau}{\tilde{\chi}}
\newcommand{\MS}{\mbox{$\overline{\rm MS}$}}
\newcommand{\MSSM}{\mbox{$\MSSM}$}

\newcommand{\tgb}{\mbox{tg$\beta$}}

 \newcommand{\zp}[3]{{Z.\ Phys.} {\bf #1} (19#2) #3}
 \newcommand{\np}[3]{{Nucl.\ Phys.} {\bf #1} (19#2)~#3}
 \newcommand{\pl}[3]{{Phys.\ Lett.} {\bf #1} (19#2) #3}
 \newcommand{\pr}[3]{{Phys.\ Rev.} {\bf #1} (19#2) #3}
 \newcommand{\prl}[3]{{Phys.\ Rev. Lett.} {\bf #1} (19#2) #3}

\begin{document}

\renewcommand{\thefootnote}{\fnsymbol{footnote}}
\setcounter{page}{0}

\begin{titlepage}

\begin{flushright}
DESY 98-207 \\
hep-ph/9812407 \\
December 1998
\end{flushright}

\vspace*{1cm}

\begin{center}
{\large \sc SUSY Particle Production at Hadron Colliders\footnote{Contribution
to the proceedings of {\it RADCOR 98}, 8--12 September 1998, Barcelona, Spain.
}}

\vspace*{1cm}

{\sc Michael Spira\footnote{Heisenberg Fellow}}

\vspace*{1cm}

{\it II.\ Institut f\"ur Theoretische Physik\footnote{Supported by
Bundesministerium f\"ur Bildung und Forschung (BMBF), Bonn, Germany,
under Contract 05~7~HH~92P~(5), and by EU Program {\it Human Capital and
Mobility} through Network {\it Physics at High Energy Colliders} under
Contract
CHRX--CT93--0357 (DG12 COMA).}, Universit\"at Hamburg, Luruper Chaussee
149, D--22761 Hamburg, Germany}

\end{center}

\vspace*{3cm}

\begin{abstract}
The determination of the full SUSY QCD corrections to the
production of squarks, gluinos and gauginos at hadron colliders is reviewed.
The NLO corrections stabilize the theoretical predictions of
the various production cross sections significantly and lead to sizeable
enhancements of the most relevant cross sections. We discuss the
phenomenological consequences of the results on present and future
experimental analyses.
\end{abstract}

\end{titlepage}

\setcounter{page}{2}

\title{SUSY Particle Production at Hadron Colliders}

\author{MICHAEL SPIRA}

\address{II.\ Institut f\"ur Theoretische Physik, Universit\"at Hamburg,
Luruper Chaussee 149, D--22761 Hamburg, Germany \\
E-mail: spira@desy.de}


\maketitle\abstracts{
The determination of the full SUSY QCD corrections to the
production of squarks, gluinos and gauginos at hadron colliders is
reviewed.
The NLO corrections stabilize the theoretical predictions of
the various production cross sections significantly and lead to sizeable
enhancements of the most relevant cross sections. We discuss the
phenomenological consequences of the results on present and future
experimental analyses.
}

\section{Introduction}
The search for Higgs bosons and
supersymmetric particles is among the most important endeavors of present and
future high energy physics.  The novel colored particles, squarks and gluinos,
and the weakly interacting gauginos can be searched for at the upgraded
Tevatron, a $p\bar p$ collider with a c.m.\ energy of 2 TeV, and the
LHC, a $pp$ collider with a c.m.\ energy of 14 TeV.  Until now the search at
the Tevatron has set the most stringent bounds on the colored SUSY particle
masses.  At the 95\% CL, gluinos have to be heavier than about 180 GeV, while
squarks with masses below about 180 GeV have been excluded for gluino masses
below $\sim 300$ GeV \cite{bounds}.  Stops, the scalar superpartners of the
top quark, have been excluded in a significant part of the MSSM parameter
space with mass less than about 80 GeV by the LEP and Tevatron experiments
\cite{bounds}.  Finally charginos with masses below about 90 GeV have been
excuded by the LEP experiments, while the present search at the Tevatron is
sensitive to chargino masses of about 60--80 GeV with a strong dependence on
the specific model \cite{bounds}. Due to the negative search at LEP2
the lightest neutralino $\tilde
\chi_1^0$ has to be heavier than about 30 GeV in the context of SUGRA
models \cite{bounds}. In the $R$-parity-conserving MSSM,
supersymmetric particles can only be produced in pairs.  All supersymmetric
particles will decay to the lightest supersymmetric particle (LSP), which is
most likely to be a neutralino, stable thanks to conserved $R$-parity.  Thus
the final signatures for the production of supersymmetric particles will
mainly be jets, charged leptons and missing transverse energy, which is
carried away by neutrinos and the invisible neutral LSP.

In Section 2 we shall summarize the
details of the calculation of the NLO QCD corrections, as described in
Refs.\ \citer{sqgl,gaunlo} for
the case of $\sq \sqb$ production.  The evaluation of the full SUSY QCD
corrections splits into two pieces, the virtual corrections, generated by
virtual particle exchanges, and the real corrections, which originate from
gluon radiation and the corresponding crossed processes with three-particle
final states.

In Section 3 we shall consider the production of squarks and gluinos except
stops \cite{sqgl}.  We assume the light-flavored squarks to be mass
degenerate, which is a
reasonable approximation for all squark flavors except stops, while the light
quarks ($u,d,s,c,b$) will be treated as massless particles.
The production of stop pairs requires the inclusion of mass splitting and
mixing effects \cite{stops} and
will be investigated in Section 4. In Section 5 we will summarize the
results for the production of charginos and neutralinos at
NLO \cite{gaunlo}. The calculation of the LO cross sections has been
performed a long time
ago \cite{lo}.  Since the [unphysical] scale dependence of the LO quantities
amounts up to about 50\%, the determination of the NLO corrections is necessary
in order to gain a reliable theoretical prediction, which can be used in
present and future experimental analyses.

\section{SUSY QCD corrections}

\subsection{Virtual corrections}
\begin{figure}[hbt]
\begin{center}
\begin{picture}(120,80)(20,10)

\Gluon(0,20)(50,20){-3}{5}
\Gluon(0,80)(50,80){3}{5}
\Gluon(50,50)(75,80){-3}{4}
\DashLine(100,20)(50,20){5}
\DashLine(50,80)(100,80){5}
\DashLine(50,20)(50,80){5}
\put(-15,78){$g$}
\put(-15,18){$g$}
\put(75,48){$g$}
\put(105,18){$\sqb$}
\put(105,78){$\sq$}

\end{picture}
\begin{picture}(120,80)(-40,10)

\Gluon(0,20)(50,20){-3}{5}
\Gluon(0,80)(50,80){-3}{5}
\Gluon(75,20)(75,80){-3}{5}
\Line(75,80)(75,20)
\ArrowLine(75,20)(50,20)
\ArrowLine(50,20)(50,80)
\ArrowLine(50,80)(75,80)
\DashLine(100,20)(75,20){5}
\DashLine(75,80)(100,80){5}
\put(-15,78){$g$}
\put(-15,18){$g$}
\put(40,48){$q$}
\put(85,48){$\gl$}
\put(105,18){$\sqb$}
\put(105,78){$\sq$}

\end{picture}  \\
\caption[]{\label{fg:virt} \it Typical diagrams of the virtual corrections.}
\end{center}
\end{figure}
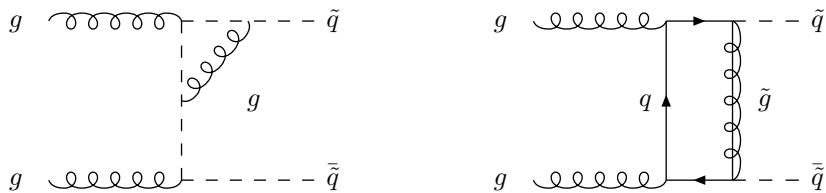
The one-loop virtual corrections are built up by gluon, gluino, quark and
squark exchange
contributions [see Fig.~\ref{fg:virt}]. They have to be contracted with the
LO matrix elements. The calculation of the one-loop contributions has been
performed in dimensional regularization, leading to the extraction of
ultraviolet, infrared and collinear singularities as poles in
$\epsilon = (4-n)/2$. For the chiral $\gamma_5$ coupling we have used the naive
scheme, which is well justified in the present analysis at the one-loop
level\footnote{We have explicitely checked that the results obtained
with a consistent $\gamma_5$ scheme are identical to the one with the naive
scheme.}.
We have explicitly checked that after summing all virtual corrections no
quadratic divergences are left over, in accordance with the general property
of supersymmetric theories. The renormalization has been performed by
identifying the squark and gluino masses with their pole masses, and defining
the strong
coupling in the $\overline{\rm MS}$ scheme including five light flavors in the
corresponding $\beta$ function. The massive particles, i.e.\ squarks, gluinos
and top quarks, have been decoupled by subtracting their contribution at
vanishing momentum transfer \cite{decouple}. In dimensional regularization,
there is a mismatch between the gluonic degrees of freedom [d.o.f. = $n-2$] and
those of the gluino [d.o.f. = $2$], so that SUSY is explicitly broken. In
order to restore SUSY in the physical observables in the massless limit, an
additional finite counter-term is required for the renormalization of the novel
$\sq \gl \bar q$ vertex \cite{count}.

\subsection{Real corrections}
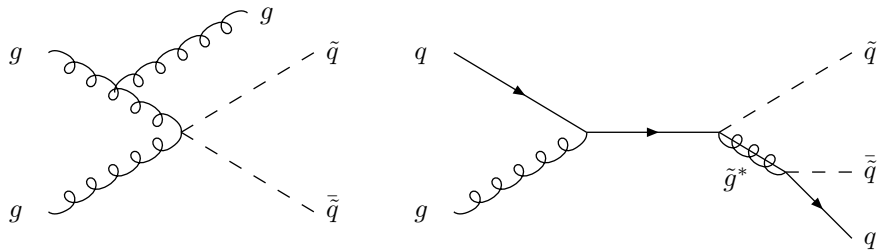
\begin{figure}[hbt]
\begin{center}
\setlength{\unitlength}{1pt}
\begin{picture}(120,80)(0,10)

\Gluon(0,20)(50,50){-3}{5}
\Gluon(0,80)(50,50){3}{5}
\Gluon(25,65)(75,95){3}{5}
\DashLine(50,50)(100,80){5}
\DashLine(100,20)(50,50){5}
\put(-15,78){$g$}
\put(-15,18){$g$}
\put(80,93){$g$}
\put(105,18){$\sqb$}
\put(105,78){$\sq$}

\end{picture}
\begin{picture}(170,80)(-30,10)

\Gluon(0,20)(50,50){-3}{5}
\ArrowLine(0,80)(50,50)
\ArrowLine(50,50)(100,50)
\DashLine(100,50)(150,80){5}
\Gluon(100,50)(125,35){-3}{3}
\Line(100,50)(125,35)
\ArrowLine(125,35)(150,10)
\DashLine(125,35)(150,35){5}
\put(-15,78){$q$}
\put(-15,18){$g$}
\put(102,30){$\gl^*$}
\put(155,78){$\sq$}
\put(155,33){$\sqb$}
\put(155,8){$q$}

\end{picture}  \\
\setlength{\unitlength}{1pt}
\caption[]{\label{fg:real} \it Typical diagrams of the real corrections.}
\end{center}
\end{figure}
The real corrections originate from the radiation of a gluon in all possible
ways and from the crossed processes by interchanging the gluon of the final
state against a light quark in the initial state.  The phase-space integration
of the final-state particles has been performed in $n=4-2\epsilon$ dimensions,
leading to the extraction of infrared and collinear singularities as poles in
$\epsilon$.  After evaluating all angular integrals and adding the virtual and
real corrections, the infrared singularities cancel.  The left-over collinear
singularities are universal and are absorbed in the renormalization of the
parton densities at NLO.  We defined the parton densities in the conventional
$\overline{\rm MS}$ scheme including five light flavors, i.e.\ the squark,
gluino and top quark contributions are not included in the mass factorization.
Finally we end up with an ultraviolet, infrared and collinear finite partonic
cross section.

However, there is an additional class of physical singularities, which have to
be regularized. In the second diagram of Fig.~\ref{fg:real} an
intermediate $\sq \gl^*$ state is produced, before the [off-shell] gluino splits
into a $q\sqb$ pair. If the gluino mass is larger than the common squark mass,
and the partonic c.m.\ energy is larger than the sum of the squark and gluino
masses, the intermediate gluino can be produced on its mass-shell. Thus the
real corrections to $\sq \sqb$ production contain a contribution of $\sq \gl$
production. The residue of this part corresponds to $\sq \gl$ production with
the subsequent gluino decay $\gl \to \sqb q$, which is already contained
in the LO cross section
of $\sq \gl$ pair production, including all final-state cascade decays.
This term has to be subtracted in order to derive a well-defined production
cross section. Analogous subtractions emerge in all reactions: if the gluino
mass is larger than the squark mass, the contributions from $\gl \to \sq \bar
q, \sqb q$ have to be subtracted, and in the reverse case the contributions of
squark decays into gluinos have to subtracted.

\section{Production of Squarks and Gluinos}
\begin{figure}[hbt]
\vspace*{-3.0cm}

\hspace*{-1.5cm}
\begin{turn}{-90}%
\epsfxsize=10cm \epsfbox{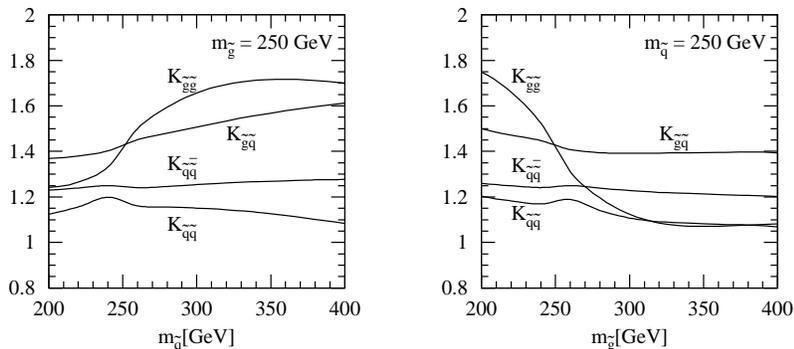}
\end{turn}
\vspace*{-3.0cm}

\caption[]{\label{fg:kfac} \it K factors of the different squark and gluino
production cross sections at the upgraded Tevatron [$\sqrt{s} = 2$ TeV].
Parton densities: CTEQ4L (LO) and CTEQ4M (NLO) with $Q=m$. 
Top mass: $m_t=175$ GeV.}
\end{figure}

\begin{figure}[hbt]
\vspace*{-2.3cm}

\hspace*{-1.5cm}
\begin{turn}{-90}%
\epsfxsize=10cm \epsfbox{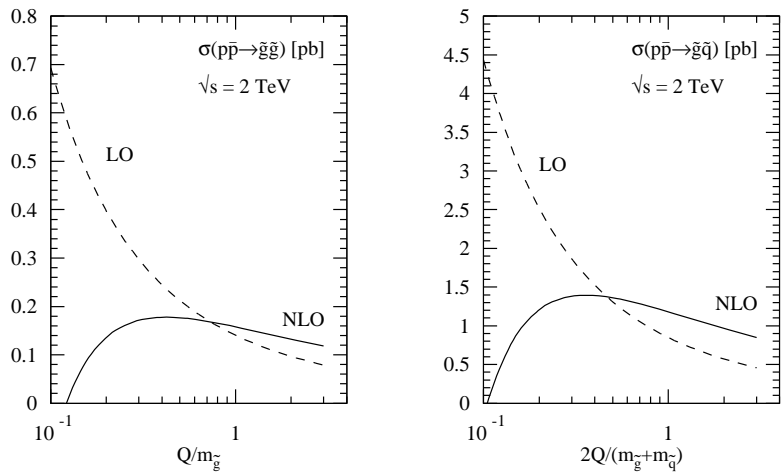}
\end{turn}
\vspace*{-2.1cm}

\caption[]{\label{fg:scale} \it Scale dependence of the
total squark and gluino production cross sections at the Tevatron in LO
and NLO. Parton densities: CTEQ4L (LO) and CTEQ4M (NLO);
mass parameters: $m_{\sq}=250$ GeV, $m_{\gl}=300$ GeV and $m_t=175$
GeV.}
\end{figure}

\begin{figure}[hbt]
\vspace*{-2.5cm}

\hspace*{-1.5cm}
\begin{turn}{-90}%
\epsfxsize=10cm \epsfbox{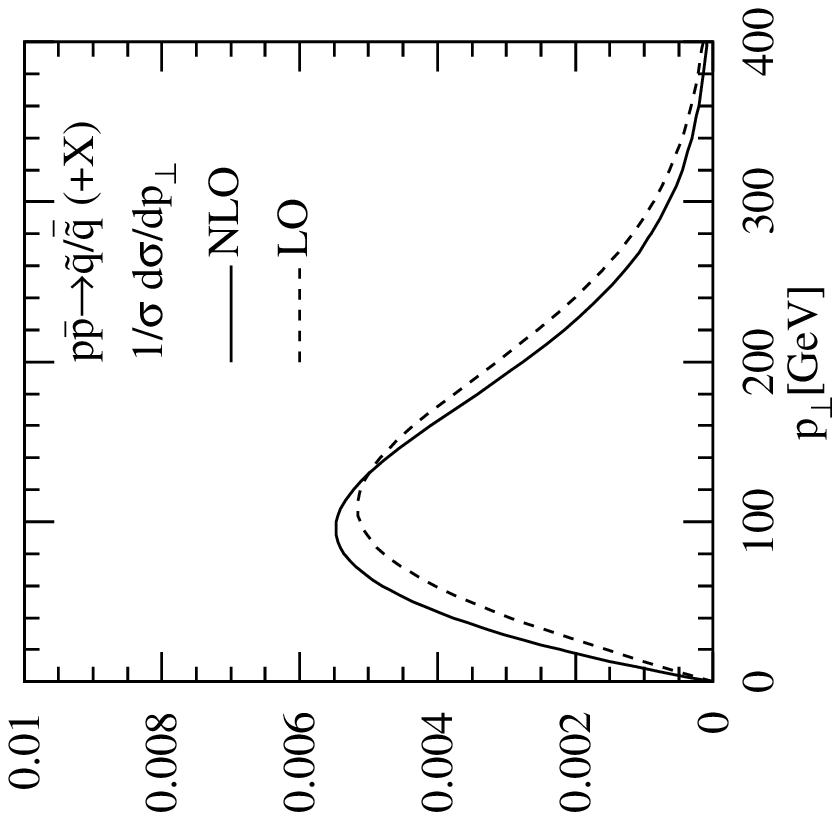}
\end{turn}

\vspace*{-10.05cm}

\hspace*{5.0cm}
\begin{turn}{-90}%
\epsfxsize=10cm \epsfbox{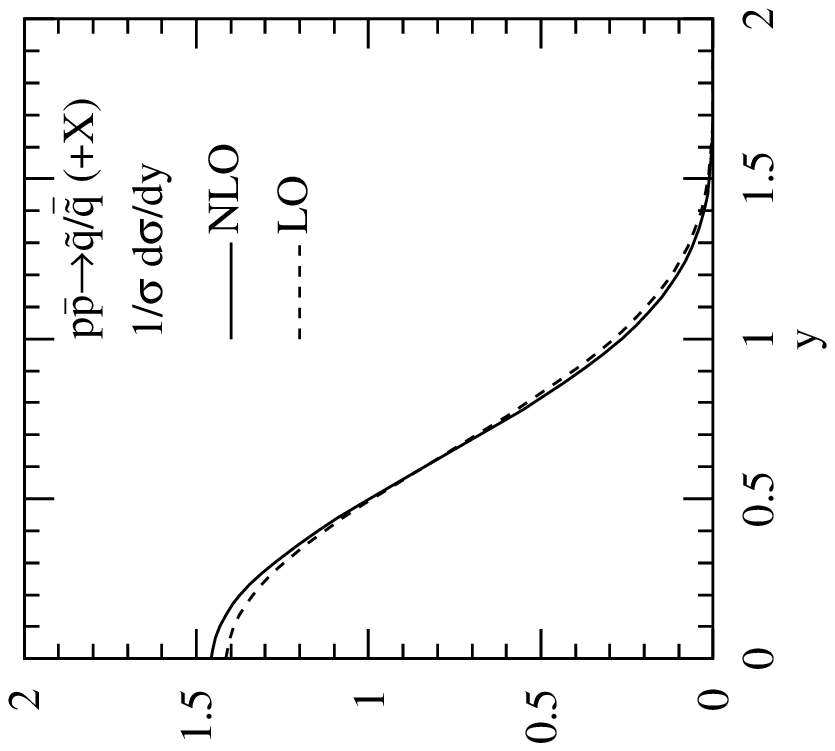}
\end{turn}
\vspace*{-2.8cm}

\caption[]{\label{fg:pty} \it Normalized transverse-momentum and rapidity
distributions of $p\bar p\to \sq \sqb + X$ at the upgraded Tevatron
[$\sqrt{s}=2$ TeV] in LO (dotted)
and NLO (solid). Parton densities: CTEQ4L (LO) and
CTEQ4M (NLO) with $Q=m$; mass parameters: $m_{\sq}=250$ GeV, $m_{\gl}=300$ GeV
and $m_t=175$ GeV.}
\end{figure}

Squarks and gluinos can be produced via $pp, p\bar p \to \sq \sqb, \sq \sq,
\sq \gl, \gl \gl$ at hadron colliders.
The hadronic squark and gluino production cross sections can be obtained from
the partonic ones by convolution with the corresponding parton densities.  We
have performed the numerical analysis for the upgraded Tevatron and the
LHC. For the
natural renormalization/factorization scale choice $Q=m$, where $m$ denotes
the average mass of the final-state SUSY particles, the SUSY QCD corrections
are large and positive, increasing the total cross sections by 10--90\%
\cite{sqgl}. This is shown in Fig.~\ref{fg:kfac}, where the K factors,
defined as the ratios of the NLO and LO cross sections, are presented as a
function of the corresponding SUSY particle mass for the Tevatron.
We have investigated the residual scale dependence in LO and NLO, which is
presented in Fig.~\ref{fg:scale}.  The inclusion of the NLO corrections
reduces the LO scale dependence by a factor 3--4 and reaches a typical level
of $\sim 15\%$, which serves as an estimate of the remaining theoretical
uncertainty.  Moreover, the dependence on different sets of parton
densities is rather weak and leads to an additional uncertainty of $\sim 15\%$.
In order to quantify the effect of the NLO corrections on the
search for squarks and gluinos at hadron colliders, we have extracted the SUSY
particle masses corresponding to several fixed values of the production cross
sections.  These masses are increased by 10--30 GeV at the Tevatron and
10--50 GeV at the LHC, thus enhancing the present and future bounds on the
squark and gluino masses significantly.
Finally we have evaluated the QCD-corrected transverse-momentum and rapidity
distributions for all different processes.  As can be inferred from
Fig.~\ref{fg:pty}, the modification of the normalized distributions in NLO
compared to LO is less than about 15\% for the transverse-momentum
distributions and much less for the rapidity distributions.  Thus it is a
sufficient approximation to rescale the LO distributions uniformly by the K
factors of the total cross sections.

\section{Stop Pair Production}
\begin{figure}[hbt]
\vspace*{-0.5cm}

\hspace*{2.0cm}
\epsfxsize=8cm \epsfbox{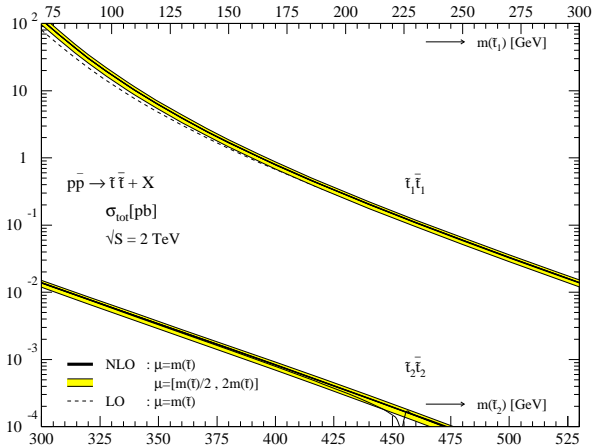}
\vspace*{-0.5cm}

\caption[]{\label{fg:stopprod} \it Production cross sections of the
light and heavy stop states at the Tevatron at LO [dashed] and NLO
[solid]. The thickness of the NLO curves represents the dependence of
the cross sections on the stop mixing angle and the squark and gluino
masses. The shaded NLO bands indicate the theoretical uncertainties due
to the scale dependence within $m_{\st}/2 < \mu < 2m_{\st}$.
Parton densities: CTEQ4L (LO) and CTEQ4M (NLO); Top mass: $m_t=175$ GeV.}
\end{figure}

At LO only pairs of $\st_1$ or pairs of $\st_2$ can be produced at hadron
colliders.  Mixed $\st_1 \st_2$ pair production is only possible at NLO and
beyond.  However, we have estimated that mixed stop pair production is
completely suppressed by several orders of magnitude and can thus safely be
neglected \cite{stops}. The evaluation of the QCD corrections proceeds along
the same lines as in the case of squarks and gluinos. The strong coupling and
the parton densities have been defined in the \MS~scheme with 5 light flavors
contributing to their scale dependences, while the stop masses are
renormalized on-shell.
\begin{figure}[hbt]
\vspace*{-3.2cm}

\hspace*{-1.5cm}
\begin{turn}{-90}%
\epsfxsize=10cm \epsfbox{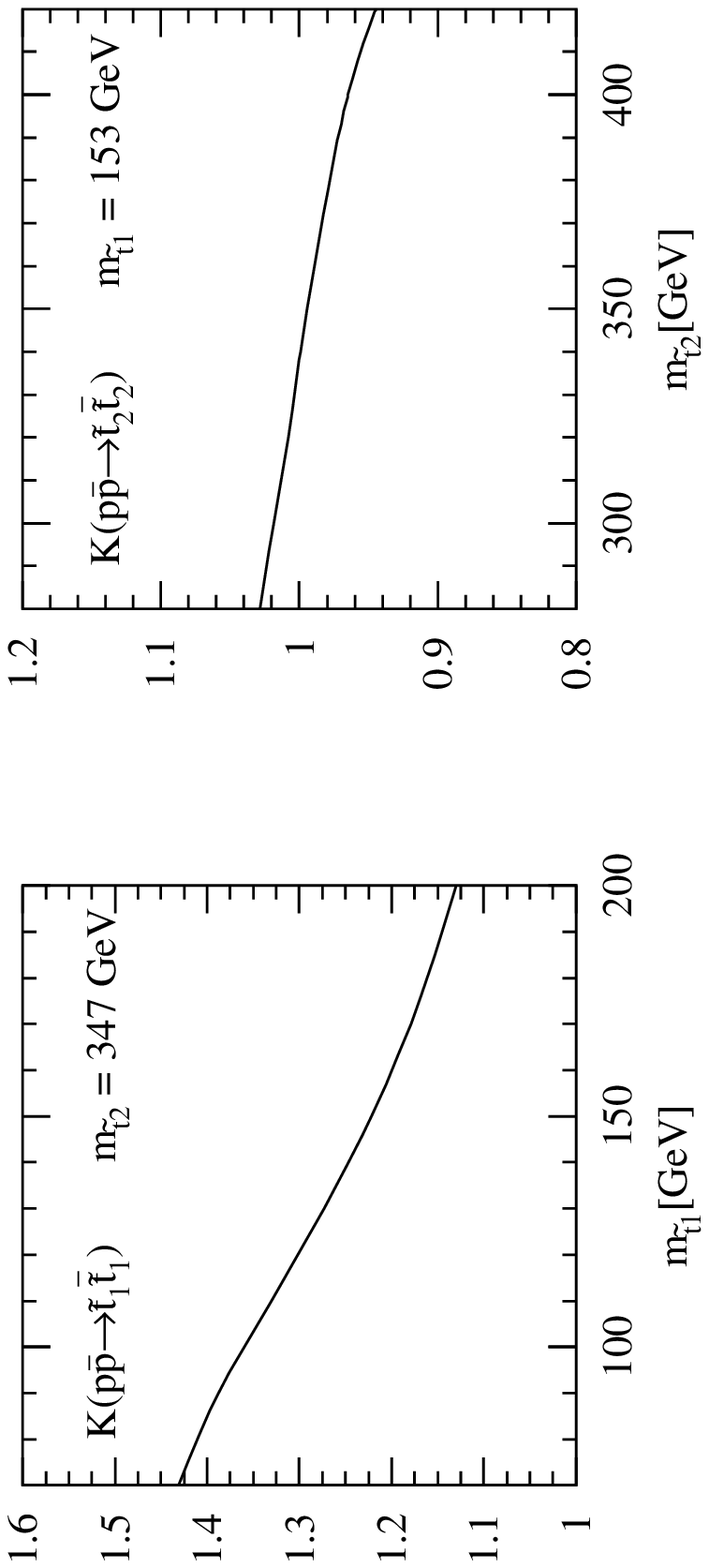}
\end{turn}
\vspace*{-3.0cm}

\caption[]{\label{fg:kst} \it K factor of the light stop
production cross sections at the upgraded Tevatron [$\sqrt{s}=2$ TeV].
Parton densities: CTEQ4L (LO)
and CTEQ4M (NLO) with $Q=m_{\st_1}$. Top mass: $m_t=175$ GeV.}
\end{figure}
The QCD corrections increase the cross sections by up to
about 40\% [see Fig.~\ref{fg:kst}] and thus lead to an increase of
the extracted stop masses from
the measurement of the total cross section.
\begin{figure}[hbt]
\vspace*{-0.0cm}

\hspace*{1.5cm}
\begin{turn}{-90}%
\epsfxsize=6cm \epsfbox{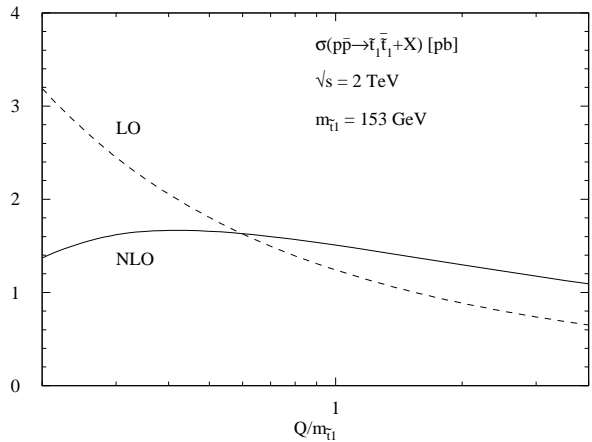}
\end{turn}
\vspace*{-0.3cm}

\caption[]{\label{fg:scst} \it Scale dependence of the
total light stop production cross sections at the Tevatron in LO
and NLO. Parton densities: CTEQ4L (LO) and CTEQ4M (NLO);
Top mass: $m_t=175$ GeV.}
\end{figure}
Moreover, as in the squark/gluino
case the scale dependence is strongly reduced [see Fig.~\ref{fg:scst}] and
yields an estimate of about
15\% of the remaining theoretical uncertainty at NLO. At NLO the virtual
corrections depend on the stop mixing angle, the squark, gluino and stop
masses of the other type. However, it turns out that these dependences are
very weak and can safely be neglected as can be inferred from
Fig.~\ref{fg:stopprod}.

\section{Chargino and Neutralino Production}
The production cross sections of charginos and neutralinos depend on several
MSSM parameters, i.e.\ $M_1, M_2, \mu$ and $\tgb$ at LO \cite{lo}. The cross
sections are sizeable for chargino/neutralino masses below about 100 GeV at
the upgraded Tevatron and less than about 200 GeV at the LHC. Due to the
strong dependence on the MSSM parameters the
extracted bounds on the chargino and neutralino masses depend on the
specific region in the MSSM parameter space \cite{bounds}. The outline of
the determination
of the QCD corrections is analogous to the previous cases of
squarks, gluinos and stops. The resonance contributions due to $gq \to \gau_i
\sq$ with $\sq \to q \gau_j$ have to be subtracted in order to avoid double
counting with the associated production of gauginos and strongly interacting
squarks and gluinos. The parton densities have been defined with 5 light
flavors contributing to their scale evolution in the \MS~scheme, while the
$t$-channel squark masses have been renormalized on-shell.
\begin{figure}[hbt]
\vspace*{-0.3cm}

\hspace*{-0.3cm}
\begin{turn}{-90}%
\epsfxsize=5cm \epsfbox{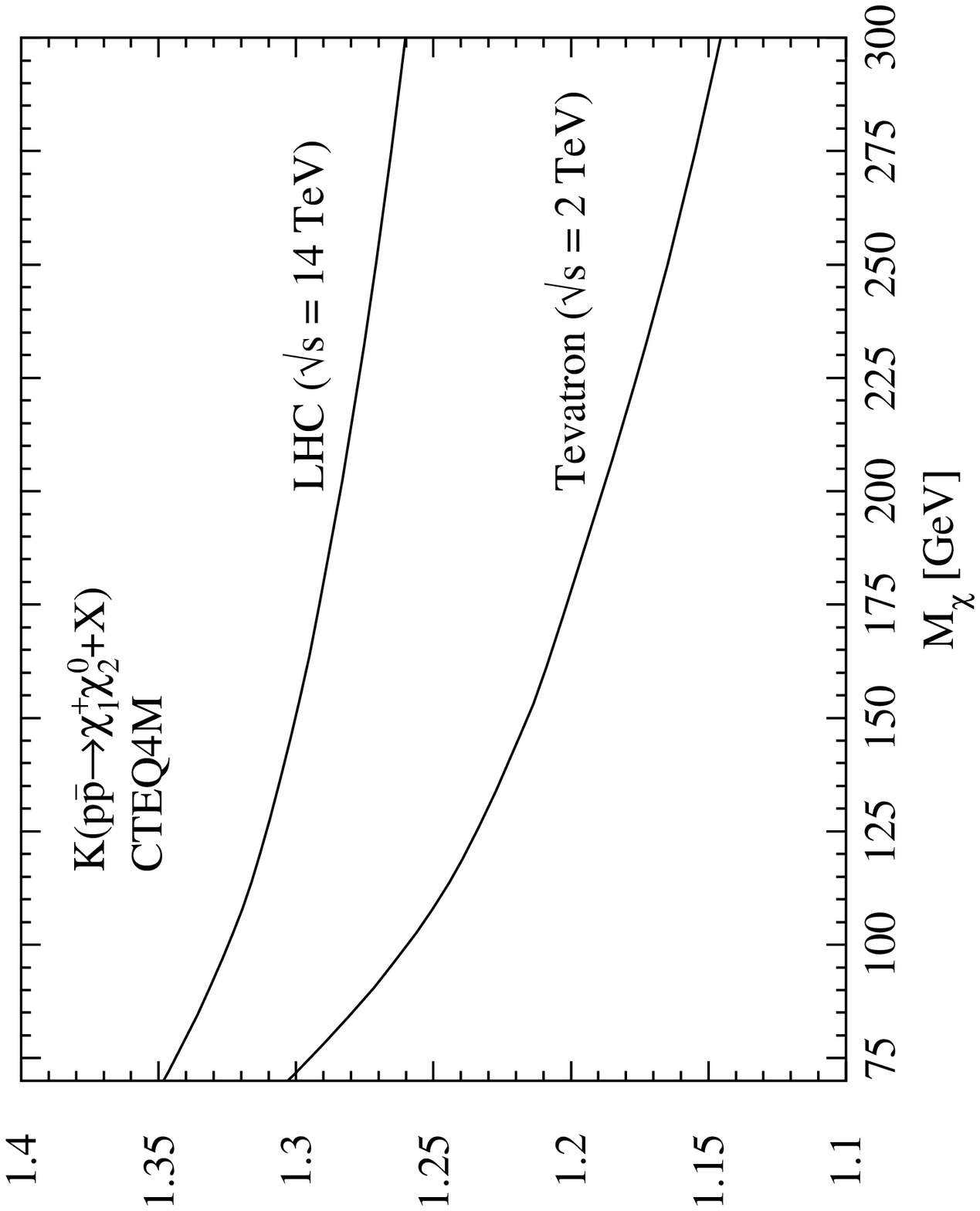}
\end{turn}
\vspace*{-5.0cm}

\hspace*{5.8cm}
\begin{turn}{-90}%
\epsfxsize=5cm \epsfbox{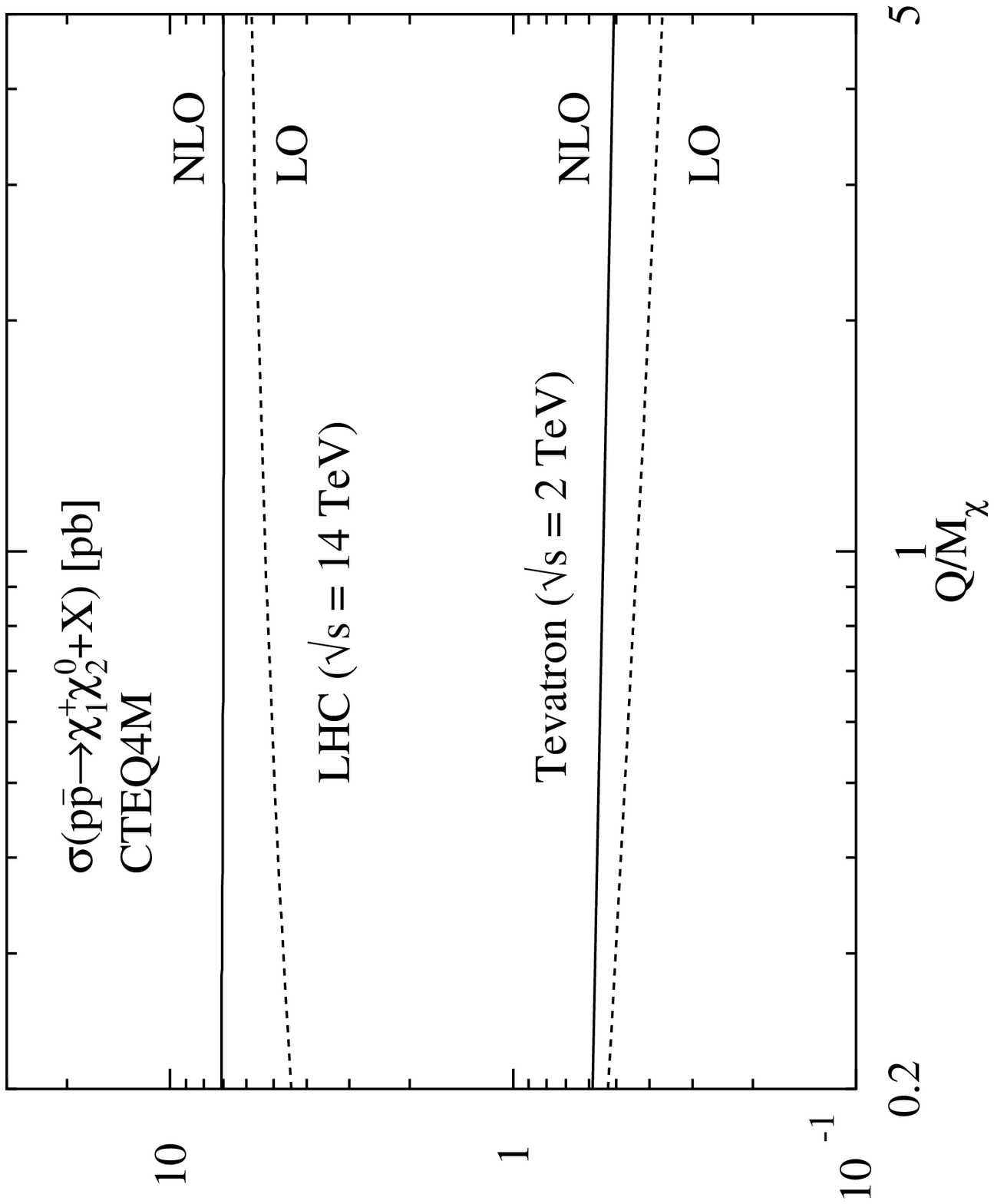}
\end{turn}
\vspace*{-0.5cm}
 
\caption[]{\label{fg:kgausc} \it K factor and scale dependence of the $\gau^+_1
\gau^0_2$
production cross section at the Tevatron and LHC. Parton densities: CTEQ4L (LO)
and CTEQ4M (NLO) with $Q=m_{\st_1}$. Top mass: $m_t=175$ GeV;
SUGRA parameters: $M_0 = 100$ GeV, $A_0 = 300$ GeV, $\tgb = 4$, $\mu > 0$.}
\end{figure}
The QCD
corrections enhance the production cross sections of charginos and
neutralinos by about 10--40\% [see Fig.~\ref{fg:kgausc}]. The LO
scale dependence is reduced to about
10\% at NLO [see Fig.~\ref{fg:kgausc}], which signalizes a significant
stabilization of the thoretical prediction for the production cross sections
\cite{gaunlo}. The dependence of the
chargino/neutralino production cross sections on the specific set of parton
densities ranges at about 15\%.

\section{Conclusions}
In this work we have reviewed the status of SUSY particle production at
hadron colliders at NLO.  Most QCD corrections to the production processes
are known, thus yielding a nearly complete theoretical status.  There are
especially large QCD corrections to the production of gluinos, which
significantly increase the extracted bounds on the gluino mass from the
negative search for these particles at the Tevatron.  In all processes, which
are known at NLO, the theoretical uncertainties are reduced to about 15\%,
which should be sufficient for the upgraded Tevatron and the LHC\footnote{The
computer code PROSPINO \cite{prospino}
for the production of squarks,
gluinos and stops at hadron colliders is available at
http://wwwcn.cern.ch/$\sim$mspira/. The NLO production of gauginos and
sleptons will be included soon.}. \\

\noindent {\bf Acknowledgements} \\
I would like to thank W.\ Beenakker, R.\ H\"opker, M.\ Kr\"amer, M.\ Klasen,
T.\ Plehn and P.M.\ Zerwas for their collaboration and the organizers of
RADCOR 98 for the pleasant atmosphere during the symposium.

\end{document}